\documentclass[aps,prl,twocolumn, titlepage,showpacs]{revtex4}
\usepackage{graphicx}
\usepackage{color}
\usepackage{dcolumn}

\usepackage{bm}

\begin{document}

\title{Frequency-dependent shear viscosity of a liquid 2D dusty plasma}

\author{Yan Feng}
\email{yan-feng@uiowa.edu}
\author{J. Goree}
\author{Bin Liu}
\affiliation{Department of Physics and Astronomy, The University
of Iowa, Iowa City, Iowa 52242, USA}

\date{\today}

\begin{abstract}

The viscoelasticity of a two-dimensional liquid strongly-coupled dusty plasma is studied experimentally, without macroscopic shear. Positions and velocities of the dust particles, measured by video microscopy, are used as the inputs to the generalized Green-Kubo relation to obtain the complex viscosity $\eta(\omega)$. The real part of $\eta(\omega)$ (which corresponds to dissipation) diminishes gradually with frequency, while the imaginary part (which corresponds to elasticity) is peaked at a frequency below the 2D dusty plasma frequency. The viscoelastic approximation is found to accurately describe the 2D experimental results for $\eta(\omega)$, yielding the Maxwell relaxation time $\tau_M = 0.10~{\rm s}$. Results for $\eta(\omega)$ are compared to 2D molecular dynamics Yukawa simulations and to a previous experiment that was performed using an oscillating macroscopic shear.

\end{abstract}

\narrowtext

\maketitle

\section {I.~INTRODUCTION}

Liquids behave as a viscous continuum in the hydrodynamic regime of large distances and long times, but in the kinetic regime where distances and times are comparable to molecular scales, liquids can also exhibit elastic effects~\cite{Hansen:86}. Viscous effects correspond to energy dissipation, while elastic effects correspond to energy storage. When they occur at the same time, as they can in a liquid in the kinetic regime, the phenomenon is called viscoelasticity~\cite{Nettleton:59}.

To characterize the viscoelasticity of a liquid quantitatively, one can generalize the concept of viscosity, so that it is dependent on both frequency $\omega$ and wave number $k$~\cite{Evans:08, Hansen:07}. It is common, however, to report a simplified quantity that depends only on frequency $\eta(\omega)$, which is a complex function:
\begin{equation}\label{sheareta}
{\eta(\omega)=\eta'(\omega) - i \eta''(\omega).}
\end{equation}
The real part $\eta'(\omega)$ corresponds to the liquid-like property of viscous dissipation, while the imaginary part $\eta''(\omega)$ corresponds to the solid-like property of elasticity~\cite{Riande:00, Smeulders:90, Donko:10}. By determining $\eta(\omega)$, one can compare the effects of energy storage and dissipation of a liquid for a particular time scale $\omega^{-1}$. A simple model of the frequency-dependent viscosity is predicted by the viscoelastic approximation~\cite{Hansen:86} which has two parameters: the Maxwell relaxation time $\tau_M$ and the static viscosity $\eta_0$, as will be discussed in Sec.~II~A. The latter quantity is the same as $\eta(\omega)$ in the hydrodynamic limit of $\omega \rightarrow 0$~\cite{Evans:08}. Aside from $\eta(\omega)$, one can also characterize viscoelasticity using the wave-number-dependent viscosity $\eta(k)$, as has been done in molecular dynamics (MD) simulations of liquids consisting of particles that interact through various potentials~\cite{Donko:10, Feng:10, Feng:11_3, Balucani:00, Hu:07} and also in a two-dimensional (2D) dusty plasma experiment~\cite{Feng:10}.

In this paper, we will investigate viscoelasticity of 2D liquids by performing experiments with video microscopy observations of particle motion in a liquid. There are several 2D physical systems that allow this kind of direct observation of individual particle motion. These include colloidal suspensions~\cite{Murray:90, Zheng:11}, granular
materials~\cite{Reis:06}, a
Wigner lattice of electrons on a liquid helium
surface~\cite{Grimes:79}, ions confined magnetically in a Penning
trap~\cite{Mitchell:99}, vortex arrays in the mixed state of type-II
superconductors~\cite{Gammel:87}, and strongly-coupled dusty plasmas
levitated in a single layer~\cite{Hartmann:11, Feng:10}.

A dusty plasma~\cite{Morfill:09, Shukla:09, Bonitz:10, Piel:10, Melzer:08, Bonitz:10_2, Shukla:02}, sometimes termed complex plasma, is a partially ionized gas containing solid micron-size particles, which we will refer to in this paper as dust particles. Due to their large length scales and slow time scales~\cite{Morfill:09}, dusty plasmas allow imaging by video microscopy and tracking of individual particle motion~\cite{Melzer:08}. Dust particles are negatively charged, with typically thousands of elementary charges. As a result, an ensemble of mutually repulsive dust particles is strongly coupled~\cite{Chu:94, Merlino:04}. Dusty plasma experiments can be performed with all the dust particles located in a single horizontal layer that is perpendicular to the ion flow; under these conditions, the collection of dust particles behaves essentially as a 2D system, with interparticle interactions that can be modeled through a repulsive Yukawa potential~\cite{Konopka:00, Lampe:00} with a screening length $\lambda_D$. Using various heating methods~\cite{Nosenko:06, Samsonov:04, Feng:08, Schella:11, Wolter:05}, the kinetic temperature~\cite{Feng:08} can be adjusted so that the suspension of dust particles behaves like a liquid or solid. In previous dusty plasma experiments, elasticity in solids~\cite{Nunomura:00} and viscosity in liquids~\cite{Juan:01, Nosenko:04, Vaulina:08, Gavrikov:10} have been studied by applying shearing stress using laser beams.

We will obtain $\eta(\omega)$ for a 2D dusty plasma using experimental data in the generalized Green-Kubo relation, as discussed in Sec.~II~C. In our literature search, we have not found any previous report, for any substance, of a similar use of the generalized Green-Kubo relation with an input of experimental data. The experimental data we use is from a 2D dusty plasma experiment that was performed earlier in our lab~\cite{Feng:11}, to maintain a 2D liquid under well-controlled steady conditions. These conditions are desirable for using the generalized Green-Kubo relation.

We also test the expressions of the viscoelastic approximation, which is a standard theory for simple liquids that has previously been tested experimentally for liquid argon and other 3D liquids~\cite{Hansen:86}, but not, to the best of our knowledge, for 2D liquids or dusty plasmas. We perform this test by fitting the equations for the viscoelastic approximation to the $\eta(\omega)$ measured in two experiments, from our lab~\cite{Feng:11} and the lab of Hartmann {\it et al.}~\cite{Hartmann:11}, both using 2D dusty plasmas.

\section {II.~FREQUENCY-DEPENDENT VISCOSITY}

In this section, we review the principles and methods for the viscoelastic approximation, and methods of obtaining the frequency-dependent viscosity $\eta(\omega)$. We will use these principles and methods in Sec.~IV.

\subsection {A. Viscoelastic approximation}
The frequency-dependent viscosity $\eta(\omega)$ characterizes how
both viscous and elastic effects vary with time scales or
frequencies. The two parts, $\eta'(\omega)$ and $\eta''(\omega)$, correspond
to viscous dissipation and elasticity, respectively. As $\omega \rightarrow 0$, it is
expected that the real part extrapolates to the usual static viscosity $\eta_0$, i.e., zero frequency viscosity.

\begin{figure}[htb]
\centering
\includegraphics{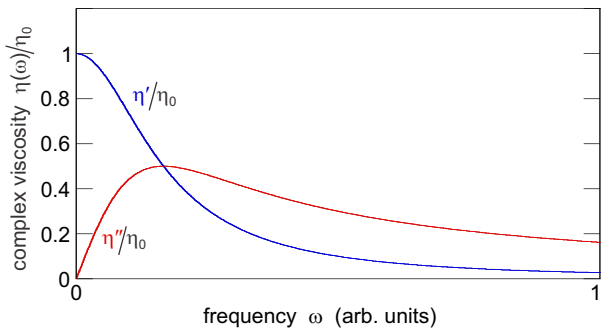}
\caption{\label{etasketch} (Color online). Sketch of the viscous and elastic parts of the
frequency-dependent viscosity, Eq.~(\ref{sheareta}), as in~\cite{Evans:08}. At low frequencies, motion is dominated by dissipation, corresponding to the real part $\eta'$, while at high frequencies, it is dominated by elasticity, corresponding to the imaginary part $\eta''$.}
\end{figure}

Figure 1 shows a sketch of a typical complex
viscosity $\eta(\omega)$ for a liquid. Both the real and imaginary
parts of $\eta(\omega)$ have been normalized by the static
viscosity $\eta_0$. As the frequency $\omega$ increases from 0,
the real part $\eta'(\omega)$ decays from unity monotonically
toward 0. The imaginary part $\eta''(\omega)$ increases
from 0 to its maximum and then decays gradually.

In the viscoelastic approximation~\cite{Hansen:86}, $\eta(\omega)$ can be expressed as
\begin{equation}\label{etatheory}
{\eta(\omega) = \frac{G_\infty}{-i\omega + 1/\tau_M},}
\end{equation}
where $G_\infty$ is an instantaneous modulus of rigidity, and $\tau_M$ is the Maxwell relaxation time, $\tau_M = \eta_0/G_\infty$. The Maxwell relaxation time $\tau_M$ characterizes the transition between viscous behavior at low frequencies ($\omega\tau_M \ll 1$) and elastic behavior at high frequencies ($\omega\tau_M \gg 1$)~\cite{Hansen:86}. The real and imaginary parts of Eq.~(\ref{etatheory}) are:
\begin{equation}\label{realetafit}
{\eta'(\omega)/\eta_0 = \frac{1}{1 + \tau_M^2 \omega^2}}
\end{equation}
for the viscous part, and
\begin{equation}\label{imaginaryetafit}
{\eta''(\omega)/\eta_0 = \frac{\tau_M \omega}{1 +
\tau_M^2 \omega^2}}
\end{equation}
for the elastic part. Equations~(\ref{realetafit}) and (\ref{imaginaryetafit}) have previously been used to fit $\eta(\omega)$ from a simulation of a 3D Yukawa liquid~\cite{Goree:11}.

In Sec.~IV, we will fit Eqs.~(\ref{realetafit}) and (\ref{imaginaryetafit}) to our experimental measurements of $\eta(\omega)$, with one free parameter, $\tau_M$. We will determine $\tau_M$ for our experiment of~\cite{Feng:11}, and compare it to values we determined from data for another experiment~\cite{Hartmann:11} and our MD simulations.

\subsection {B. Rheometry to measure $\eta(\omega)$}

A rheometer is an instrument that imposes {\it a macroscopic shear} to measure viscosity. A boundary contacting the fluid is rotated, causing a flow with a velocity gradient. The resulting shearing stress can be measured to determine the viscosity $\eta$, using its hydrodynamical definition~\cite{Pal:00}. This can be done either with a steady or oscillating rotation, to find $\eta_0$ or $\eta(\omega)$, respectively. The imaginary part of $\eta(\omega)$ is determined by measuring a phase shift (i.e., delay) between the applied modulation and the resulting shearing stress.

Dusty plasmas are not suited for use with rheometers, however, because the dust particles do not contact any solid surface. Therefore, experimenters must determine viscosity some other way, such as applying a shearing stress using laser beams~\cite{Nosenko:04, Hartmann:11}, or by using the Green-Kubo relation~\cite{Feng:11} as we will do in this paper.

\subsection {C. Generalized Green-Kubo relation to measure $\eta(\omega)$}
The generalized Green-Kubo relation can be used to obtain
the frequency-dependent viscosity. Previously it was used in
simulations~\cite{Donko:10}. It can be used for conditions {\it without macroscopic shear}. Its inputs are the positions, velocities and potential energies of individual particles as they move in their thermal motion. The frequency-dependent viscosity $\eta(\omega)$ is calculated in three steps.

First, the shearing stress $P_{xy}(t)$ is calculated as
\begin{equation}\label{SS}
{P_{xy}(t)=
\sum_{i=1}^N\left[mv_{ix}v_{iy}-\frac{1}{2}\sum_{j\not=i}^N\frac{x_{ij}y_{ij}}{r_{ij}}\frac{\partial
\Phi(r_{ij})}{\partial r_{ij}}\right].}
\end{equation}
Here, $i$ and $j$ indicate different particles, $N$ is the total
number of  particles of mass $m$, $\mathbf{r}_{i} = (x_i,y_i)$ is
the position of particle $i$, $r_{ij}=|\mathbf{r}_i-\mathbf{r}_j|$, $x_{ij}=x_i-x_j$, $y_{ij}=y_i-y_j$, and $\Phi(r_{ij})$ is the
interparticle potential energy.

Second, the shearing stress
autocorrelation function (SACF) is calculated as
\begin{equation}\label{SACF}
{C_{\eta}(t)= \langle P_{xy}(t)P_{xy}(0) \rangle.}
\end{equation}
For the experiment reported here, despite the efforts that were made, there was still a small nonzero average velocity in the analyzed region, which results in a nonzero time-averaged $P_{xy}$. To solve this problem, as in~\cite{Feng:11}, we subtract the time-average value from $P_{xy} (t)$, so that we use only the fluctuating portion to calculate the SACF.

Third, the frequency-dependent viscosity $\eta(\omega)$ can be calculated from the Laplace-Fourier transformation of $C_{\eta}(t)$, i.e., the generalized Green-Kubo relation:
\begin{equation}\label{eta}
{\eta (\omega)=\frac{1}{A k_B T}\int^\infty_0 C_{\eta}(t) e^{i \omega t} dt,}
\end{equation}
where $A$ is the area of the 2D system and $T$ is its temperature. Equation~(\ref{eta}) can be derived by combining Eqs. (20) and (21) of ~\cite{Evans:81} for a zero wavenumber, and it has been used previously for a 3D Yukawa simulation~\cite{Donko:10}. The units of $\eta$ are ${\rm kg\,s^{-1}}$ for a 2D liquid (unlike the viscosity of a 3D liquid, which has units of ${\rm kg\,m^{-1}\,s^{-1}}$). By specifying $\omega = 0$, Eq.~(\ref{eta}) is the same as the usual
Green-Kubo relation for the static viscosity $\eta_0$~\cite{Feng:11}. Unlike the shearing stress in
Eq.~(\ref{SS}), $\eta(\omega)$ in Eq.~(\ref{eta}) is independent of the system size.

The required inputs to calculate $P_{xy}(t)$ are time series of
positions, velocities, and potential energies of each particle.
For experiments that rely on video microscopy, however, only
positions and velocities are usually recorded, not $\Phi$.
For the experiment reported here, the potential energy cannot be measured directly, so we calculate it from other measurements by assuming a model potential. We use the Yukawa potential, which has been shown experimentally~\cite{Konopka:00} and theoretically~\cite{Lampe:00} to describe the conditions in a plane perpendicular to the ion flow at a distance of about $\lambda_D$, as is the case for the single layer of dust particles in our experiment~\cite{Feng:11}.

To obtain the frequency-dependent viscosity $\eta(\omega)$ using Eq.~(\ref{eta}) with an input of experimental data, it is necessary to replace the infinity upper limit with a finite time. In Sec.~IV, we will
replace the upper limit of the integral in Eq.~(\ref{eta}) with the time that $C_\eta(t)$ crosses zero for at least
two consecutive data points in the time series, as in~\cite{Feng:11}. This choice for the upper limit determines the smallest value of $\omega$ at which we can measure $\eta(\omega)$. As in Eqs.~(\ref{realetafit}) and (\ref{imaginaryetafit}), we normalize $\eta(\omega)$ by the static viscosity $\eta_0$, which we calculate using Eq.~(\ref{eta}) with $\omega = 0$, as in~\cite{Feng:11}.

\subsection {D. Review of $\eta(\omega)$ in dusty plasmas}
The viscoelasticity of dusty plasmas has been the subject of several theoretical studies~\cite{Murillo:00, Murillo:00_1, Veeresha:10, Upadhyaya:10, Ghosh:11, Banerjee:10}. It has also been studied using 3D Yukawa MD simulations~\cite{Donko:10}. Early experimental studies~\cite{Ratynskaia:06, Chan:07} were followed by measurements of the wave-number-dependent viscosity $\eta(k)$ ~\cite{Feng:10} and the frequency-dependent viscosity $\eta(\omega)$~\cite{Hartmann:11}. In the latter experiment~\cite{Hartmann:11}, a macroscopic oscillatory shear was applied to a 2D dusty plasma using a time-modulated laser manipulation.

The generalized Green-Kubo relation Eq.~(\ref{eta}) that we use in this paper is based on the fluctuating shearing stress associated with the thermal motion of the dust particles, without any macroscopic shear. Equation~(\ref{eta}) has been tested using 3D simulations~\cite{Donko:10} where it was shown that it yields $\eta(\omega)$ consistent with results from a simulation using a different method, with an oscillatory shear. Our use of Eq.~(\ref{eta}) is different because we use experimental data, and we do this for a system that is 2D instead of 3D.

In this paper, we use the {\it generalized} Green-Kubo relation with an input of experimental data to obtain the complex viscosity $\eta(\omega)$. The experiment, which was reported in~\cite{Feng:11}, is reviewed briefly in Sec.~III. Our calculation differs from that of~\cite{Feng:11} because the Green-Kubo relation that we use, Eq.~(\ref{eta}), is the generalized form, so that we obtain the frequency-dependent viscosity instead of the static viscosity $\eta_0$. In Sec.~IV we present our results for $\eta(\omega)$ and $\tau_M$, and we compare to the previous experimental results of Hartmann {\it et al.}~\cite{Hartmann:11}, who used an applied oscillatory shear. In Sec.~V we compare our experimental results to 2D MD simulations, with and without the effects of gas friction.

\section {III.~EXPERIMENT}

In this section, we briefly review the experiment of~\cite{Feng:11}, which provides the data we will use in Sec.~IV. Dust particles, which were $8.1~{\rm \mu m}$ diameter polymer microspheres, were electrically levitated as a 2D horizontal layer in an argon rf plasma. Using a top-view camera operated at $250~{\rm frame/s}$, we tracked $\approx 2100$ dust particles, yielding time series of their positions and velocities~\cite{Feng:07, Feng:11_2}. Four runs, each lasting $20.2~{\rm s}$, were performed with laser heating so that the dust particles behaved collectively as a liquid. In this liquid, the dust particle kinetic temperature was nearly uniform, with a value of $T = 2.5 \times 10^4~{\rm K}$. The ``kinetic temperature'' describes the motion of a dust particle, not the temperature of the polymer material within the dust particle. We calculate the kinetic temperature from the mean-square velocity fluctuation as in~\cite{Feng:08}.

The layer of dust particles had the following parameters. The interparticle distance, characterized by the Wigner-Seitz radius~\cite{Kalman:04} $a=0.35~{\rm mm}$, corresponded to an areal mass density of $\rho = 1.1 \times 10^{-6}~{\rm kg/m^2}$. A moving dust particle experienced a drag force due to the neutral gas characterized by a friction coefficient $\nu_f=2.4~{\rm s^{-1}}$. The charge on a dust particle was $Q/e = -6000$, the 2D dust plasma frequency~\cite{Kalman:04} was $\omega_{pd} = 30~{\rm s^{-1}}$, the coupling parameter was $\Gamma = (Q^2 /4 \pi \epsilon_0 a)/ (k_B T) = 68$, and the screening parameter was $\kappa = a/\lambda_D = 0.5$. The values of $Q$ and $\kappa$ have uncertainties estimated to be $\pm 10~\%$ and $\pm 20~\%$, respectively. Further details of this experiment can be found in~\cite{Feng:11}.

The result for the shearing stress autocorrelation function (SACF) for one experimental run is shown in Fig.~2. We calculate the shearing stress autocorrelation function (SACF) using Eq.~(\ref{SACF}), after obtaining $P_{xy}(t)$ using Eq.~(\ref{SS}). The SACF generally has a trend of decay, starting with its maximum value at $t = 0$. The frequency content within this decay is important in the determination of $\eta(\omega)$.

\begin{figure}[htb]
\centering
\includegraphics{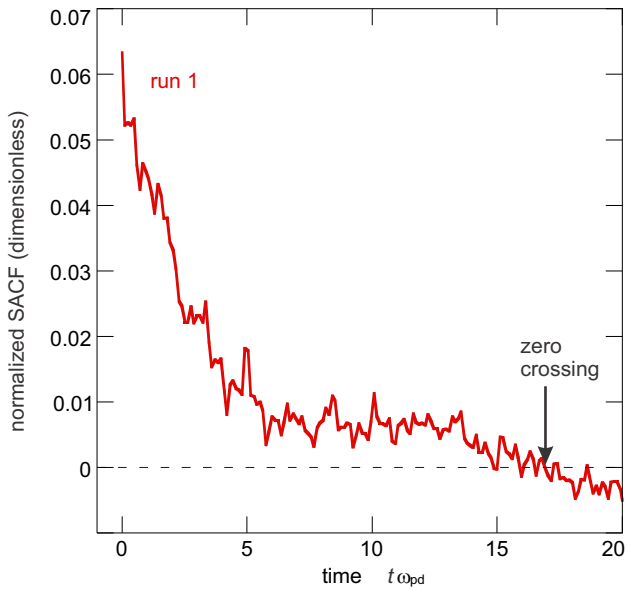}
\caption{\label{SACFFig} (Color online). The shearing
stress autocorrelation function (SACF), i.e., $C_\eta$, calculated using an input of experimental data from one
experimental run. The vertical axis, the SACF, is
normalized by $(Ak_BT\rho a^2\omega_{pd}^2)$, while the horizontal axis, time, is
normalized by $\omega_{pd}^{-1}$.}
\end{figure}

In Sec.~IV we will analyze data from four runs to obtain the frequency dependent viscosity $\eta(\omega)$ using Eq.~(\ref{eta}). This result for $\eta(\omega)$ will have an uncertainty of about $13\%$ due to the uncertainties in $Q$ and $\kappa$.

\section {IV.~RESULTS}

\subsection {A. Frequency-dependent viscosity $\eta(\omega)$}

In Fig.~3 we present our chief result: the frequency dependence of the complex viscosity, $\eta(\omega)$. These data points are obtained using the generalized Green-Kubo relation, Eq.~(\ref{eta}), after the SACF is calculated, for each run. We also obtain the zero-frequency static viscosity $\eta_0$ for that run by using Eq.~(\ref{eta}) with $\omega = 0$. Repeating for all four runs, we obtain the data for $\eta(\omega)$ shown in Fig.~3.

\begin{figure}[htb]
\centering
\includegraphics{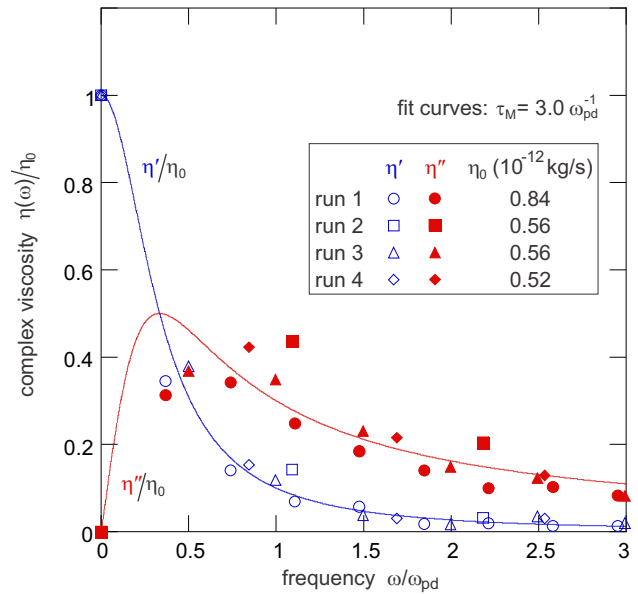}
\caption{\label{etaFig} (Color online). The frequency-dependent viscosity $\eta(\omega)$ obtained using an input of experimental data in Eq.~(\ref{eta}). These results are shown as data points for four experimental runs. The smooth curves are the result of a simultaneous fit using the viscoelastic approximation, Eqs.~(\ref{realetafit}) and (\ref{imaginaryetafit}). This fit yields the Maxwell relaxation time, $\tau_M = 0.10~{\rm s} = 3.0\,\omega_{pd}^{-1}$. The vertical axis is normalized by the static viscosity $\eta_0$ for each run.}
\end{figure}

The trends seen in our results for $\eta'(\omega)$ and $\eta''(\omega)$ in Fig.~3 for our 2D liquid are generally the same as for 3D liquids. The real part $\eta'(\omega)$ (which corresponds to dissipation) decays with increasing frequency. The imaginary part $\eta''(\omega)$ starts at zero for $\omega = 0$, then has a peak followed by a decay at large $\omega$.

To allow a comparison to other experiments and simulations, our results for the frequency-dependent viscosity $\eta(\omega)$ in Fig.~3 are presented in normalized units. Viscosity is normalized by the static viscosity $\eta_0$, which was determined separately for each run. Frequency is normalized as $\omega / \omega_{pd}$, where $\omega_{pd}$ is the 2D dust plasma frequency. The scatter in our results for $\eta(\omega)$ in Fig.~3 arises from noise in the SACF, which is believed to be statistical noise due to the finite amount of data used in a calculation for a single experimental run.

\subsection {B. Test of the viscoelastic approximation}

\subsubsection {1. Without macroscopic shear}

We now test the viscoelastic approximation, using the results of the frequency-dependent viscosity obtained from the experiment of~\cite{Feng:11}, which had no macroscopic shear. We fit the expressions for the viscoelastic approximation, Eqs.~(\ref{realetafit}) and (\ref{imaginaryetafit}), to our experimental results for the frequency-dependent viscosity data, yielding the smooth curves in Fig.~3. In performing this fit, we combined the data for all four runs in Fig.~3. This fit has a single free parameter, which is the Maxwell relaxation time $\tau_M$.

We find $\tau_M  = 0.10~{\rm s}$, which in dimensionless units is $3.0\,\omega_{pd}^{-1}$. The physical significance of the Maxwell relaxation time is that elastic behavior dominates at shorter times $t \ll \tau_M$ while viscous dissipation dominates at longer times $t \gg \tau_M$.

\subsubsection {2. With macroscopic shear}

\begin{figure}[htb]
\centering
\includegraphics{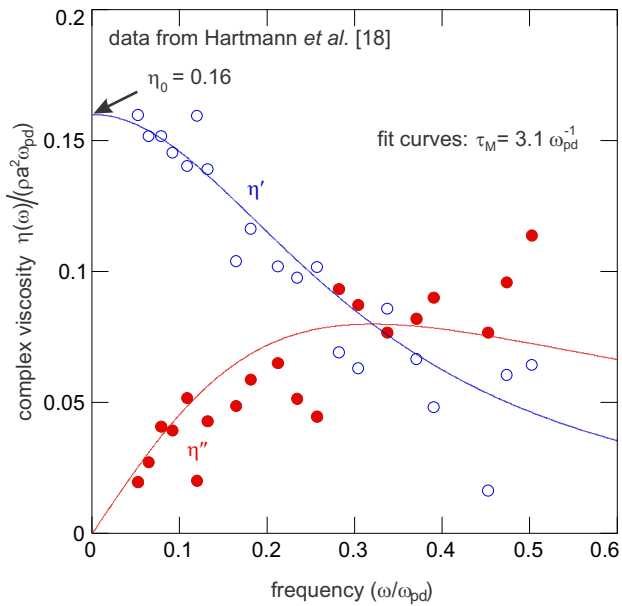}
\caption{\label{etaH} (Color online). The frequency-dependent viscosity $\eta(\omega)$ from the experiment of Hartmann {\it et al.}~\cite{Hartmann:11}, which was performed using an oscillatory shear. Here, the viscosity has been normalized by $\rho a^2 \omega_{pd}$, differently from Figs.~3 and 5. The smooth curves are our fit using the viscoelastic approximation, yielding $\tau_M = 3.1\,\omega_{pd}^{-1}$ and $\eta_0 = 0.16$. Another time scale of interest corresponds to gas friction, which in this experiment is quantified by $\nu_f = 0.04\,\omega_{pd}$.}
\end{figure}

For comparision, we also test the viscoelastic approximation, this time using data from the experiment of Hartmann {\it et al.}~\cite{Hartmann:11}, which had a macroscopic oscillating shear. That experiment was similar to ours in several ways. Both experiments used a single layer suspension of melamine-formaldehyde microspheres immersed in a radio-frequency argon plasma, and the dust particles were tracked using video microscopy.

The two experiments used different laser manipulation schemes and different approaches to determine $\eta(\omega)$. We used a uniform laser heating and the Green-Kubo approach, which requires that the dusty plasma have steady uniform conditions, while Hartmann {\it et al.} applied an oscillatory shear at an adjustable frequency and obtained the viscosity as the ratio of the shearing stress $P_{xy}$ and the applied shear rate. The Green-Kubo approach tends to generate data most easily at high frequencies, because we replace the infinite time limit in Eq.~(\ref{eta}) with a finite time, as mentioned in Sec.~II. On the other hand, the data reported by Hartmann {\it et al.} were mainly for lower frequencies. Thus, the two methods are complementary to each other.

Some of the experimental conditions were different. In~\cite{Feng:11} the dust-particle diameter was nearly twice as large as in~\cite{Hartmann:11}, and the gas pressure was slightly higher in~\cite{Feng:11}. The screening length $\lambda_D$ was approximately twice as large in~\cite{Feng:11}, even though the particle spacing was nearly the same, so that the screening parameters $\kappa$ were different, 0.5 for~\cite{Feng:11} but 1.2 for~\cite{Hartmann:11}.

We find that the viscoelastic approximation is capable of describing both experiments. This is seen in Fig.~4, where we see that the fit of the viscoelastic approximation Eqs.~(\ref{realetafit}) and (\ref{imaginaryetafit}) passes through the data points of Hartmann {\it et al.}~\cite{Hartmann:11}.  The values of the fit parameters $\tau_M$ and $\eta_0$ for the two experiments are not expected to match, since the experimental conditions were different. Nevertheless, we find that both experiments had the same normalized static viscosity~\cite{Feng:11} $\eta_0 / \rho a^2 \omega_{pd}= 0.16 $. We also find that both experiments had nearly the same normalized relaxation time: $\tau_M = 3.0\,\omega_{pd}^{-1}$ and $\tau_M = 3.1\,\omega_{pd}^{-1}$, for the experiments of~\cite{Feng:11} and~\cite{Hartmann:11}, respectively.

\section {V.~SIMULATIONS}

\begin{figure}[htb]
\centering
\includegraphics{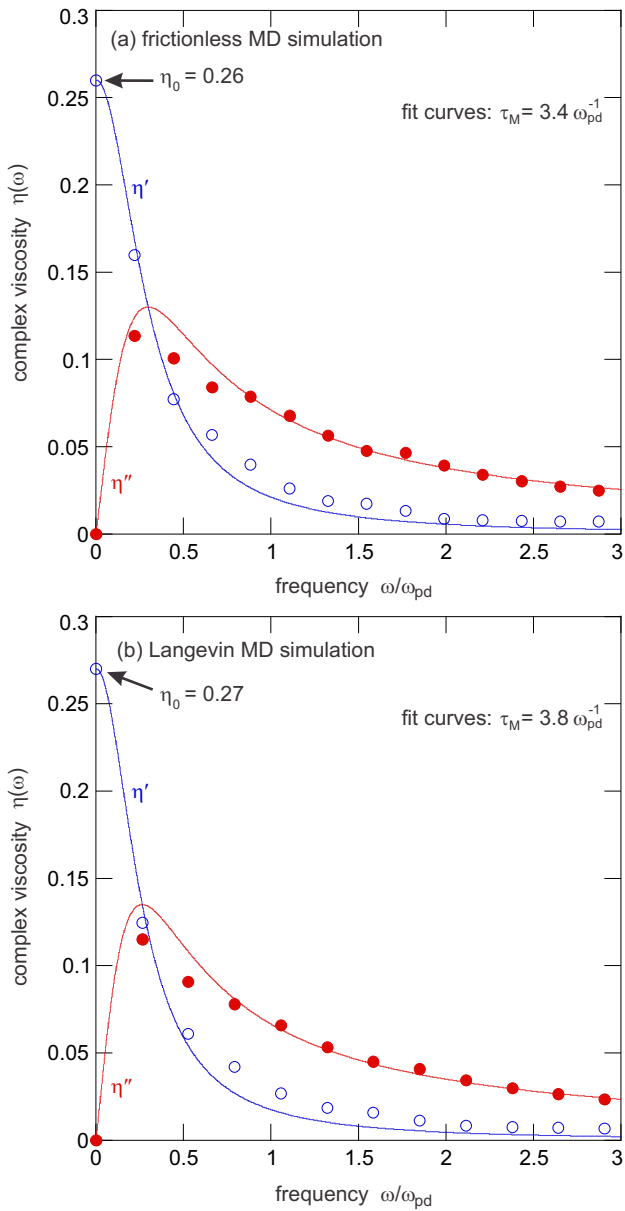}
\caption{\label{etasim} (Color online). Simulation results for a 2D Yukawa liquid. The frequency-dependent viscosity is accurately described by the viscoelastic approximation, shown by the smooth curves. The simulation used the same values of $\Gamma$ and $\kappa$ as for our experiment of~\cite{Feng:11}, and it was performed two ways, using the (a) frictionless and (b) Langevin methods.}
\end{figure}

To demonstrate that the viscoelasticity that we observed for our dusty plasma experiment is the result of interactions of the dust particles, we performed MD simulations. These simulations used only simple physics, with an integration of the equations of motion of dust particles interacting through a binary Yukawa potential. Using the particle positions, velocities and potentials recorded in the simulations, we calculated $\eta'(\omega)$ and $\eta''(\omega)$. We then fit Eqs.~(\ref{realetafit}) and (\ref{imaginaryetafit}) to these results to obtain $\tau_M$.

Two simulation methods were used: the frictionless MD and the Langevin MD methods, as described in~\cite{Feng:11_3}. Both used a binary interparticle interaction
with a Yukawa pair potential for particles that were constrained to move in a 2D plane. For both
simulations, we used $N = 4096$ particles in a
rectangular box with periodic boundary conditions. Trajectories
$\mathbf{r}_i(t)$ were found by integrating equations of motion for
all particles in MD simulations~\cite{Feng:11_3}, with the integration
time step of $0.019~\omega_{pd}^{-1}$. The simulation data trajectories were
recorded for a duration $68~000~\omega_{pd}^{-1}$ after a
steady state was reached. This duration was more than $100\times$ longer than one experimental run, so that the simulation has better statistics. The dimensionless parameters $\Gamma$ and
$\kappa$ were chosen to match
the experimental values of $\Gamma=68$ and
$\kappa=0.5$~\cite{Feng:10}. In our Langevin MD simulation, we chose the gas friction to be the same as in the experiment, $\nu_f = 0.08~\omega_{pd}$. Ewald summation~\cite{ Hamaguchi:96} was not required because our 2D simulation box is large enough compared to the screening length~\cite{Liu:05}.

Simulation results for $\eta'(\omega)$ and $\eta''(\omega)$ are
shown in Fig.~5. They exhibit the same trends as in the experimental results. The Maxwell relaxation time, $\tau_M$, is found to be $3.4\,\omega_{pd}^{-1}$ for the frictionless MD simulation, and $3.8\,\omega_{pd}^{-1}$ for the Langevin MD simulation. These two values are both close to the experimental value $\tau_M = 3.0\,\omega_{pd}^{-1}$, for the same $\Gamma$ and $\kappa$. This agreement between the experiment and these simulations demonstrates
that the viscoelasticity observed in the experiment is mainly due to binary interactions among dust particles.

\section {VI. SUMMARY}

We experimentally determined the frequency-dependent viscosity $\eta(\omega)$ for a 2D dusty plasma, using the generalized Green-Kubo relation. This measurement relies on the thermal motion of individual dust particles for experimental conditions without any macroscopic shear. We found that the real and imaginary parts of $\eta(\omega)$ show agreement with the viscoelastic approximation, so that we were also able to determine the value of the Maxwell relaxation time, $\tau_M = 0.10~{\rm s} = 3.0\,\omega_{pd}^{-1}$ for our experimental conditions.

We performed 2D MD simulations, using the same parameters as in the experiment, and we found that the simulations predict $\eta(\omega)$ and $\tau_M$ that agree with the experiment. This comparison to simulation indicates that the viscoelastic effects in the 2D dusty plasma, as quantified by $\eta(\omega)$, are mainly due to the binary interparticle interactions among the dust particles.

The generalized Green-Kubo relation that we used for conditions without externally applied shear yielded data for $\eta(\omega)$ mainly for higher frequencies. In contrast, Hartmann {\it et al.}~\cite{Hartmann:11}, who used an applied oscillatory shear, reported results for lower frequencies. We find that the viscoelastic approximation is capable of describing both experiments.

This work was supported by NSF and NASA.

\end{document}